\documentclass[useAMS,usenatbib,usegraphicx]{mn2e}
\title{Distance to the SNR CTB 109/AXP 1E 2259+586 by HI absorption and self-absorption} 

\author[W.W. Tian, D.A. Leahy, D. Li]{W. W. Tian$^{1,2}$ 
\thanks{E-mail: wtian@ucalgary.ca; tww@bao.ac.cn},  
D.A. Leahy$^{2}$, D. Li$^{3}$\\
$^{1}$National Astronomical Observatories, CAS, Beijing 100012, China\\
$^{2}$Department of Physics $\&$ Astronomy, University of Calgary, Calgary,
Alberta T2N 1N4, Canada\\
$^{3}$Jet Propulsion Laboratory, California Institute of Technology, Pasadena, CA 91109} 

\begin{document}
\date{Accepted Feb. 2, 2010; Received 2009; in original form 2009 XX}
\pagerange{\pageref{firstpage}--\pageref{lastpage}} \pubyear{}

\maketitle

\label{firstpage}
\begin{abstract}
We suggest a revised distance to the supernova remnant (SNR) G109.1-1.0 (CTB 109) and its associated anomalous X-ray pulsar (AXP) 1E 2259+586 by analyzing 21cm HI-line and $^{12}$CO-line spectra of CTB 109, HII region Sh 152, and the adjacent molecular cloud complex. CTB 109 has been established to be interacting with a large molecular cloud (recession velocity at $v=-$55 km s$^{-1}$). The highest radial velocities of absorption features towards CTB 109 ($-$56 km s$^{-1}$) and Sh 152 ($-$65 km s$^{-1}$) are larger than the recombination line velocity ($-$50 km s$^{-1}$) of Sh 152 demonstrating the velocity reversal within the Perseus arm. The molecular cloud has cold HI column density large enough to produce HI self-absorption (HISA) and HI narrow self-absorption (HINSA) if it was at the near side of the velocity reversal. Absence of both HISA and HINSA indicates that the cloud is at the far side of the velocity reversal within the Perseus Arm, so we obtain a distance for CTB 109 of 4$\pm$0.8 kpc. The new distance still leads to a normal explosion energy for CTB 109/AXP 1E 2259+586.
  
\end{abstract}


\section{Introduction}

\begin{figure*}
\vspace{40mm}
\begin{picture}(50,50)
\put(-210,-30){\includegraphics{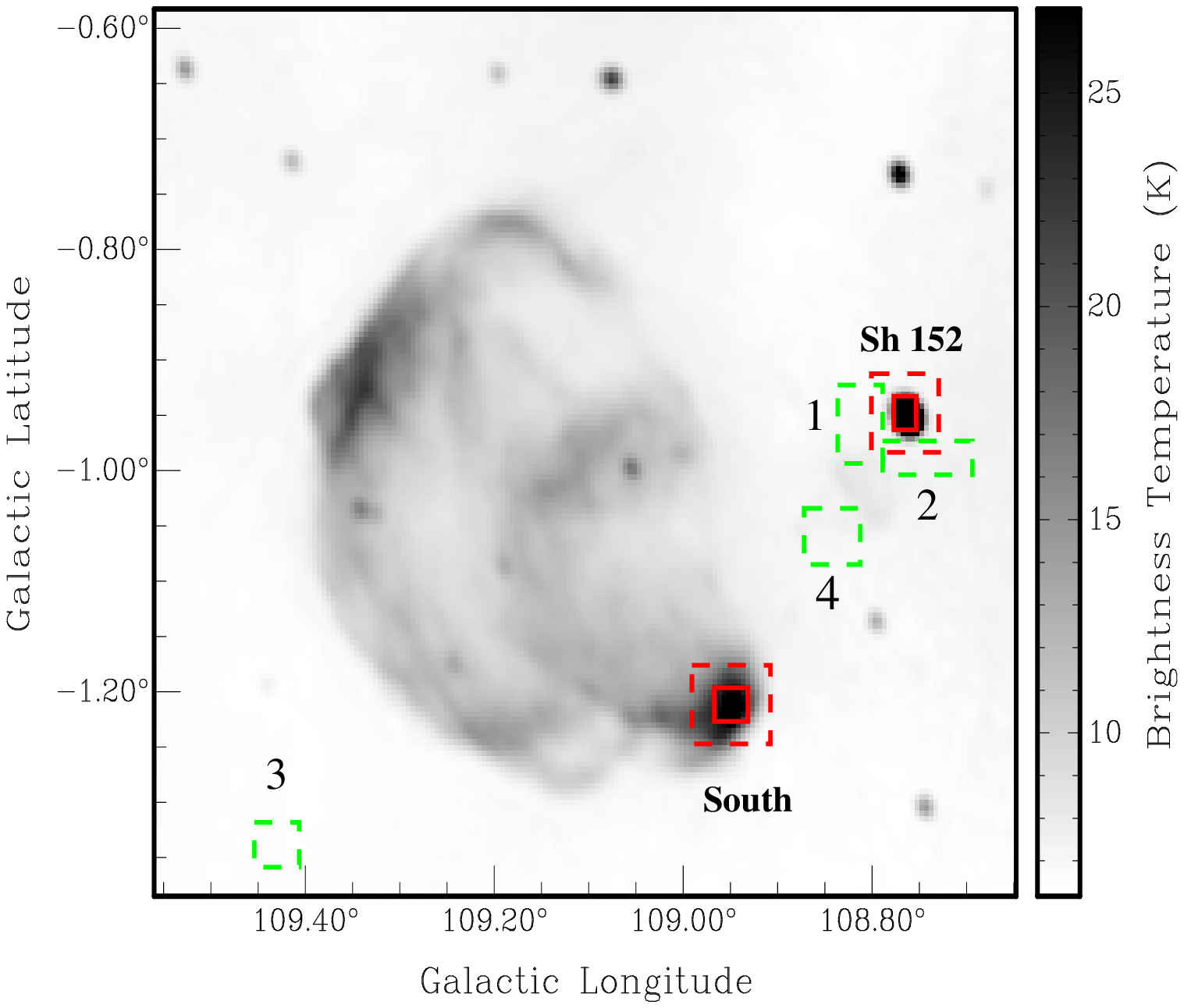}}
\put(50,-63){\includegraphics{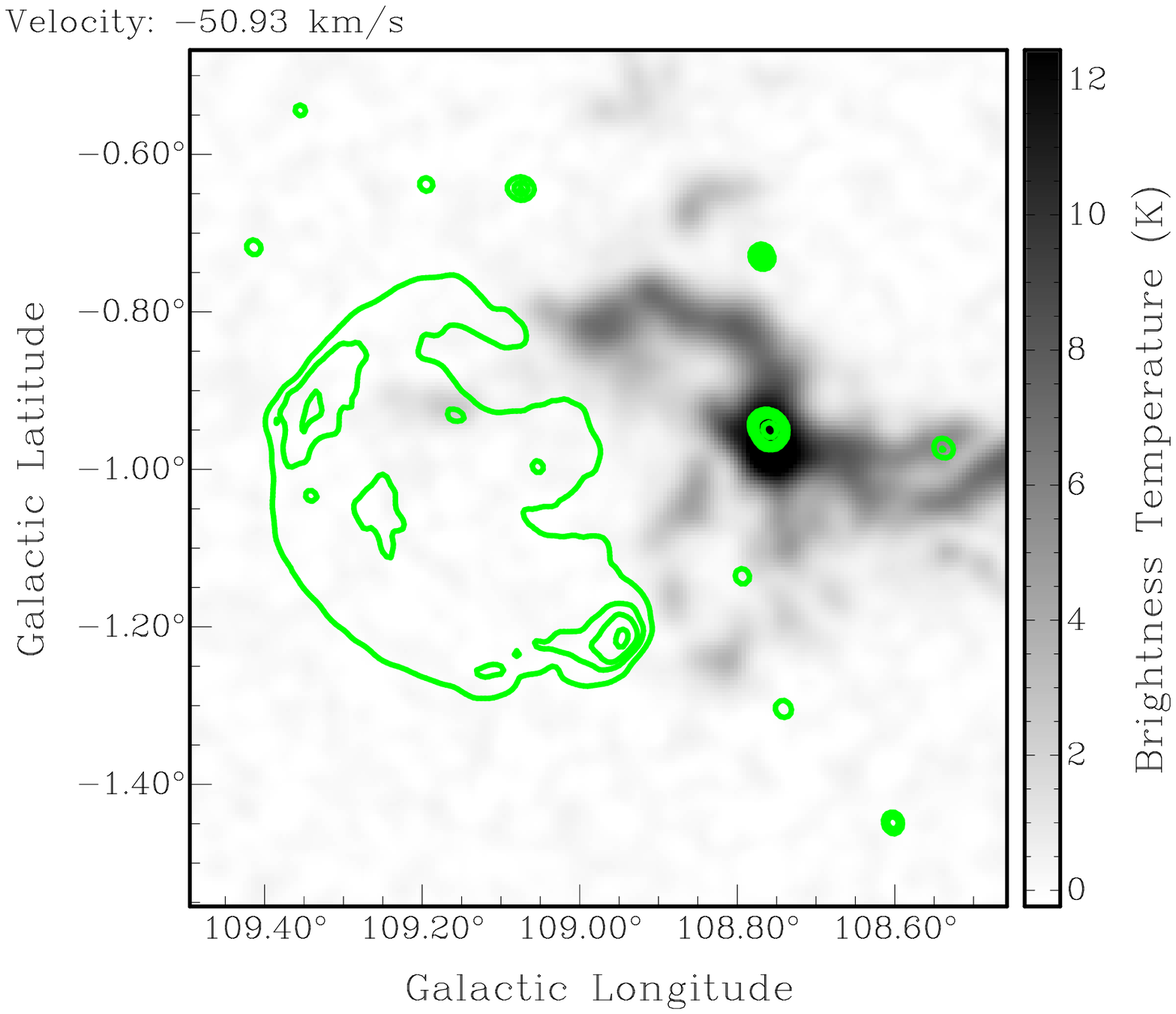}}
\end{picture}
\caption{Left: 1420 MHz continuum image of CTB 109 and the near HII region Sh 152. Right:  CO emission at $-$51 km s$^{-1}$ with contours (8, 15, 21, 29, 100 K) of the continuum emission. The left panel shows boxes (both green and red) used for extraction of HI spectra (see text) and centered at [$l=108.80^\circ$,$b=-0.96^\circ$] for box 1,  [$108.76^\circ$,$-0.98^\circ$] for box 2, [$109.43^\circ$,$-1.36^\circ$] for box 3, [$108.76^\circ$,$-0.96^\circ$], [$108.84^\circ$,$-1.02^\circ$] for box 4, [$108.77^\circ$,$-0.95^\circ$] for Sh 152 and [$108.93^\circ$,$-1.22^\circ$] for CTB 109, respectively}
\end{figure*}

The Supernova Remnant (SNR) CTB 109 hosts an anomalous X-ray
pulsar (AXP) 1E 2259+586 with a superstrong magnetic field, i.e. a magnetar. 1E 2259+586 emitted a Soft Gamma-ray Repeater-like outburst in 2002 \citep{Kaset03}, so CTB 109 is an especially interesting target for further study. CTB 109 is believed to strongly interact with the large molecular cloud ($l$=108.77$^\circ$, $b$=-0.95$^\circ$) adjacent to its west side, based on its semi-circular shape in both X-rays and radio \citep{Tatet90, Coeet89}. CTB 109 has an extended X-ray bright region in the interior, which is thermal in origin \citep*{Rhoet97} and has no morphological correlation with AXP 1E 2259+586 \citep{Patet01}. The lobe likely originates from an interaction of the SNR shock with a cloud at $-$55 km s$^{-1}$ \citep*{Saset04, Saset06}.

Distance measurement to a SNR/AXP system is extremely important but usually quite difficult. Many efforts have previously been aimed at distance determination of SNR/pulsar systems \citep{Camet06, Gotet09, Leaet08, McCet05, Rivet10, Suet09, Tiaet06, Tiaet08a, Weiet08}.  \citet[K02 hereafter]{Kotet02} analyzed HI absorption and emission spectra of CTB 109 and near compact radio sources, and estimated a distance of 3.0 kpc to CTB 109.

On the other hand, \citet*[DvK06 hereafter]{Duret06} used 
the red clump stars method to measure distances to AXPs.
This method allows deriving the function of reddening versus distance in any given line of sight, based on field stars over a relatively small area.  DvK06 excluded a distance of 3 kpc for AXP 1E 2259+586 based on its high extinction, instead preferring a much larger value of 6 kpc. 

The conflict between these two distances
motivates us to re-analyze  distance evidence for CTB 109 using 1420 MHz continuum, 21 cm HI-line and $^{12}$CO data. 
We provide a revised distance to CTB 109 by employing improved HI spectral analysis methods, and by analyzing possible HI Narrow Self-Absorption (HINSA) and HI Self-Absorption (HISA) from the adjacent molecular cloud complex. Based on similar considerations, the methods have recently been used to measure kinematic distances to Galactic molecular clouds \citep{Romet09, Krcet08}. 
We compare our new results to the previous conflicting distance estimates of K02 and DvK06.

\section{Results and Analysis}

 We show the 1420 MHz continuum image (left) and the CO image at  $v=-$51 km s$^{-1}$ (right) in Figure 1, both centered on CTB 109.
As suggested by previous studies, there is a good correlation between CTB 109 and CO features in the velocity range $-$50 to $-$57 km s$^{-1}$ (see Fig. 4 of K02), but no respective convincing HI features in the velocity range associated with CTB 109. 

\begin{figure} 
\vspace{105mm} 
\begin{picture}(80,80)
\put(-10,180){\includegraphics{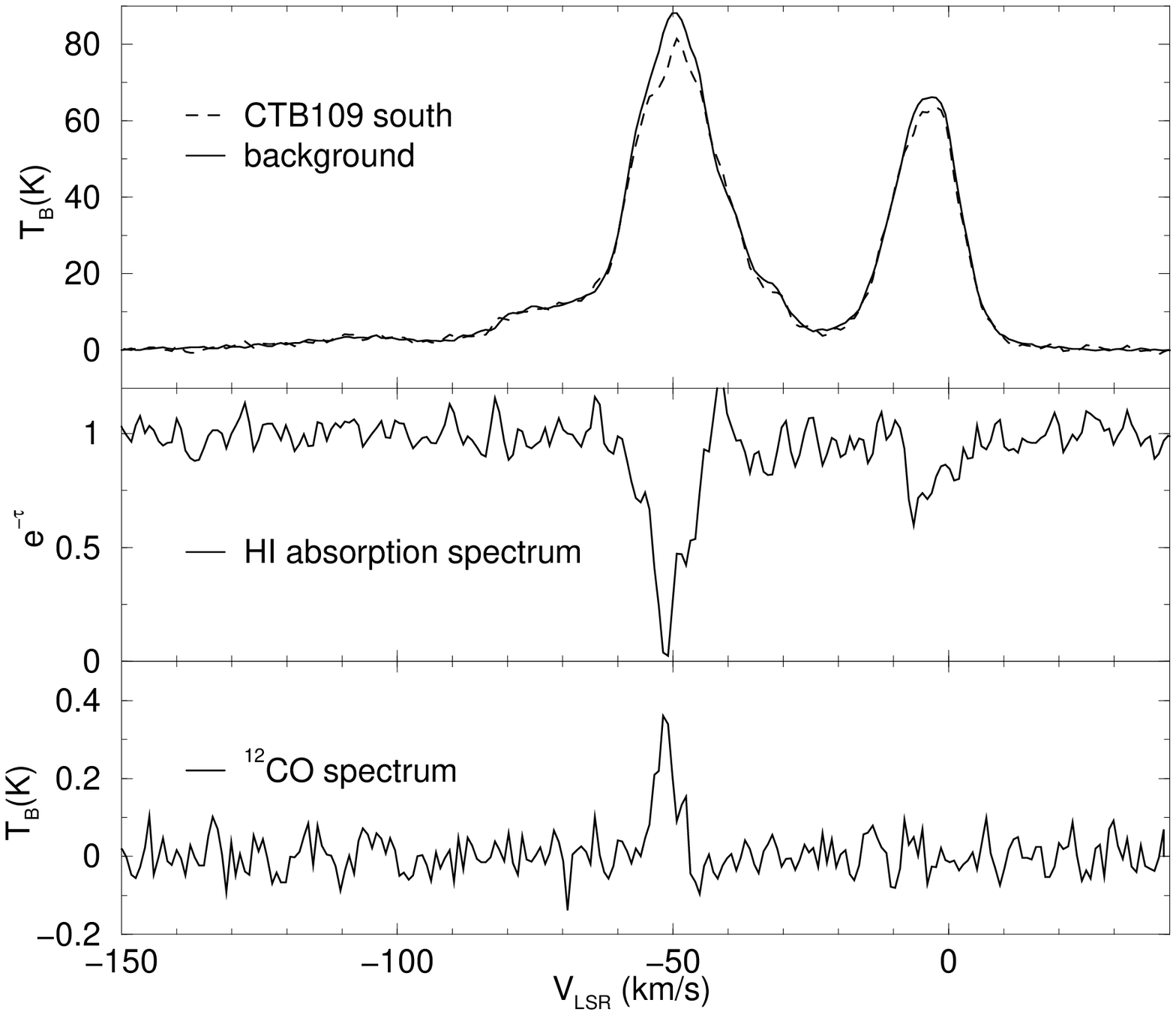}}
\put(-10,-15){\includegraphics{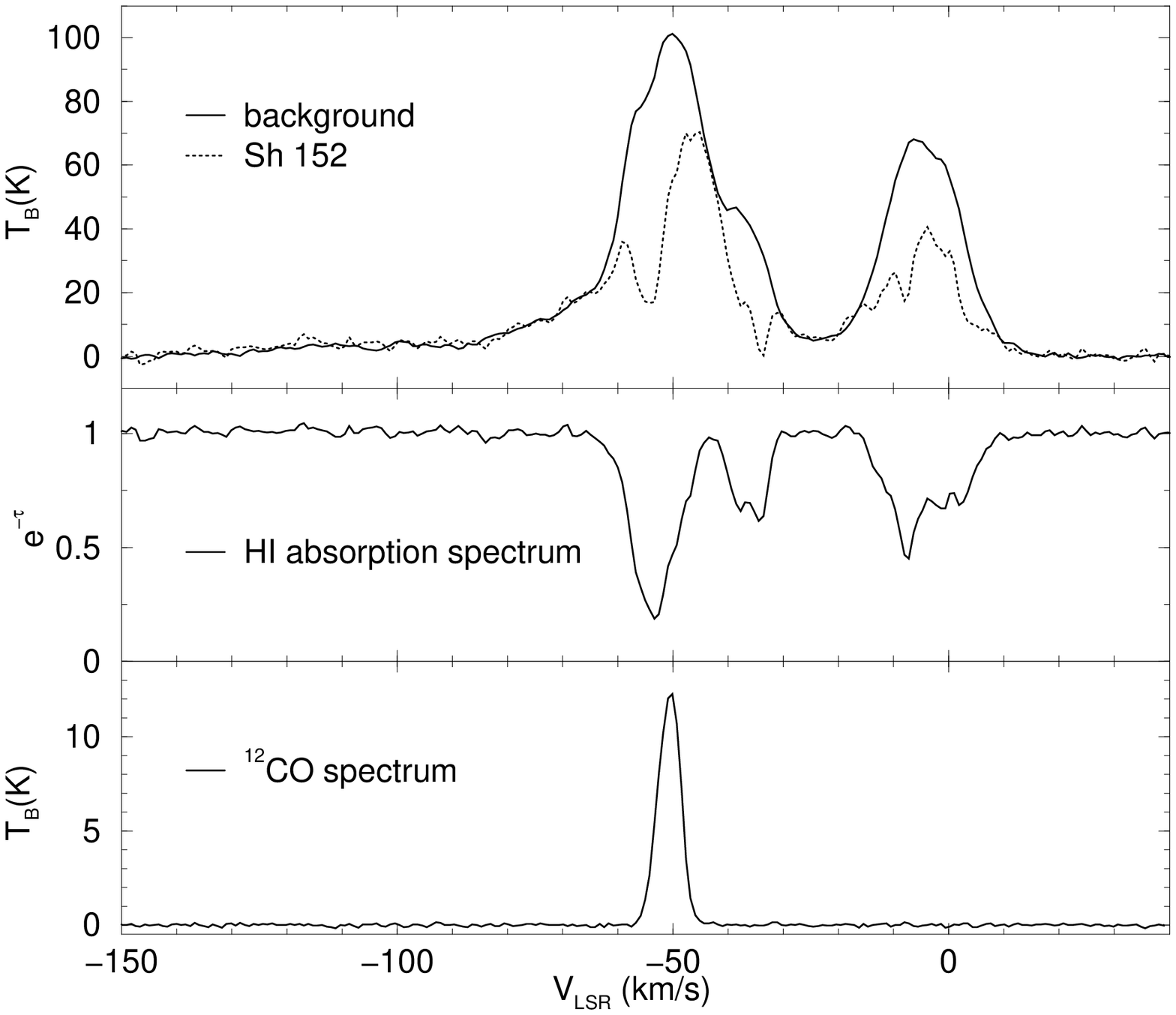}}
\end{picture}
\caption{HI and $^{12}$CO spectra of CTB 109 and Sh 152}
\end{figure} 

\subsection{HI absorption}
We construct HI emission and absorption spectra for CTB 109 and for Sh 152.  
Since CTB 109 is extended rather than a point source, the standard formula for HI absorption spectra does not apply. The continuum emission from CTB 109 extends into our background region because the background region is chosen to be near the continuum peak. This minimizes the difference in the HI distribution along the two lines of sight (source and background, see \citet*{Tiaet08b}. 
  The resulting difference in brightness temperature is given by: $\Delta T$ = $T_{off}(\it{v})$ - $T_{on}(\it{v})$ = ($T^{c}_{s}$-$T^{c}_{bg}$)(1-$e^{-\tau_{c}(\it{v})}$).  
Here $T_{on}(\it{v})$ and  $T_{off}(\it{v})$ are the average HI brightness temperatures of spectra from a selected area on a strong continuum emission region of the source and an adjacent background region which contains weaker continuum emission. $T^{c}_{s}$ and $T^{c}_{bg}$ are the average continuum brightness temperatures for the same regions respectively. $\tau_{c}(\it{v})$ is the HI optical depth from the continuum source to the observer.

 The regions used for source and background spectra are shown by the solid-line and dashed-line boxes (red; the background area excludes the source area) in the left panel of Fig. 1. Fig. 2 gives the HI and $^{12}$CO spectra.  
 The spectrum from the south filament of CTB 109 (top in Fig. 2) shows a highest HI absorption velocity of $-$56 km s$^{-1}$. The HI absorption spectrum from Sh 152 (botton) is the same shape as that given by K02, and shows a highest absorption velocity of $-$65 km s$^{-1}$. 

Sh 152 sits at the peak of a large high brightness-temperature CO molecular cloud at velocities of $-$48 to$ -$56 km s$^{-1}$ (see Fig. 4 of K02, and bottom in Fig. 2). There is clear HI absorption against Sh 152 in the same velocity range. This means cold HI gas surrounding the CO cloud 
is responsible for the HI absorption and is located in front of Sh 152. 

The CO spectrum of CTB 109 shows narrow CO emission at $-$48 to $-$54 km s$^{-1}$ in the direction of the southern filament. The fact that the highest radial velocities of HI absorption features towards both CTB 109 ($-$56 km s$^{-1}$) and Sh 152 ($-$65 km s$^{-1}$) are larger than the recombination line velocity ($-$50 km s$^{-1}$, \citet{Loc89} of Sh 152 confirms the existence of a velocity reversal within the Perseus arm. This reversal explains why the HI gas at higher radial velocity is in front of Sh 152 which is at lower radial velocity. 

Based on Fig. 2, the HI column densities integrating over all velocity in front of CTB 109 and Sh 152  are N$^{total}_{HI}$ $\sim$ 4 and 6 $\times$10$^{19}$$T_{s}$ cm$^{-2}$ respectively (using N$_{HI}$= 1.82$\times$10$^{18}\tau\Delta{v}T_{s}$, \citet*{Dicet90}). For CTB 109, the column density of HI from $-$48 to $-$54 km s$^{-1}$ is $\sim$3$\times$10$^{19}$$T_{s}$ cm$^{-2}$ (3$\times$10$^{21}$ cm$^{-2}$ if $T_{s}$$\sim$ 100K).

\subsection{HI self-absorption and narrow self-absorption} 
HI self-absorption (HISA) is generally produced by temperature fluctuations in the cold neutral medium (CNM). HISA features in the near Perseus arm have been widely revealed in the CGPS \citep{Gibet05} due to the above-mentioned velocity reversal caused by spiral density waves \citep{Rob72}. A particular kind of HI self-absorption, referred to as HI narrow self-absorption (HINSA, \citet*{Liet03}) is produced by cold HI mixed with molecular clouds and cooled by its molecular environment. HINSA is shown to share the spatial and kinematic structure of cold molecular clouds traced by CO emission lines \citep{Krcet08}.  HINSA can only be detected in specific molecular cloud regions which have narrower line width (1 to 2 km s$^{-1}$) than background HI emission from CNM components.  HINSA bears special relevance to this work because the originating location of the cold HI can be constrained by studying CO clouds at the same velocity given their co-location.

The CO image and spectrum in the direction of Sh 152 show a large molecular cloud complex in the velocity range of $-$48 to $-$56 km s$^{-1}$. 
Part of the CO cloud interacting with CTB 109 has been believed to be a major factor causing CTB 109 to have an asymmetric semi-circular shape structure, i.e. a relatively bright eastern part, well-defined in radio and X-ray images, and no emission in the west. Interaction of the SNR shock with the part of this molecular cloud complex at $-55$ km s$^{-1}$ has been observed \citep{Saset06}. HINSA could be produced if these CO clouds are in the front of warm HI background. 
 
\begin{figure}
\vspace{140mm}
\begin{picture}(80,80)
\put(0,350){\includegraphics{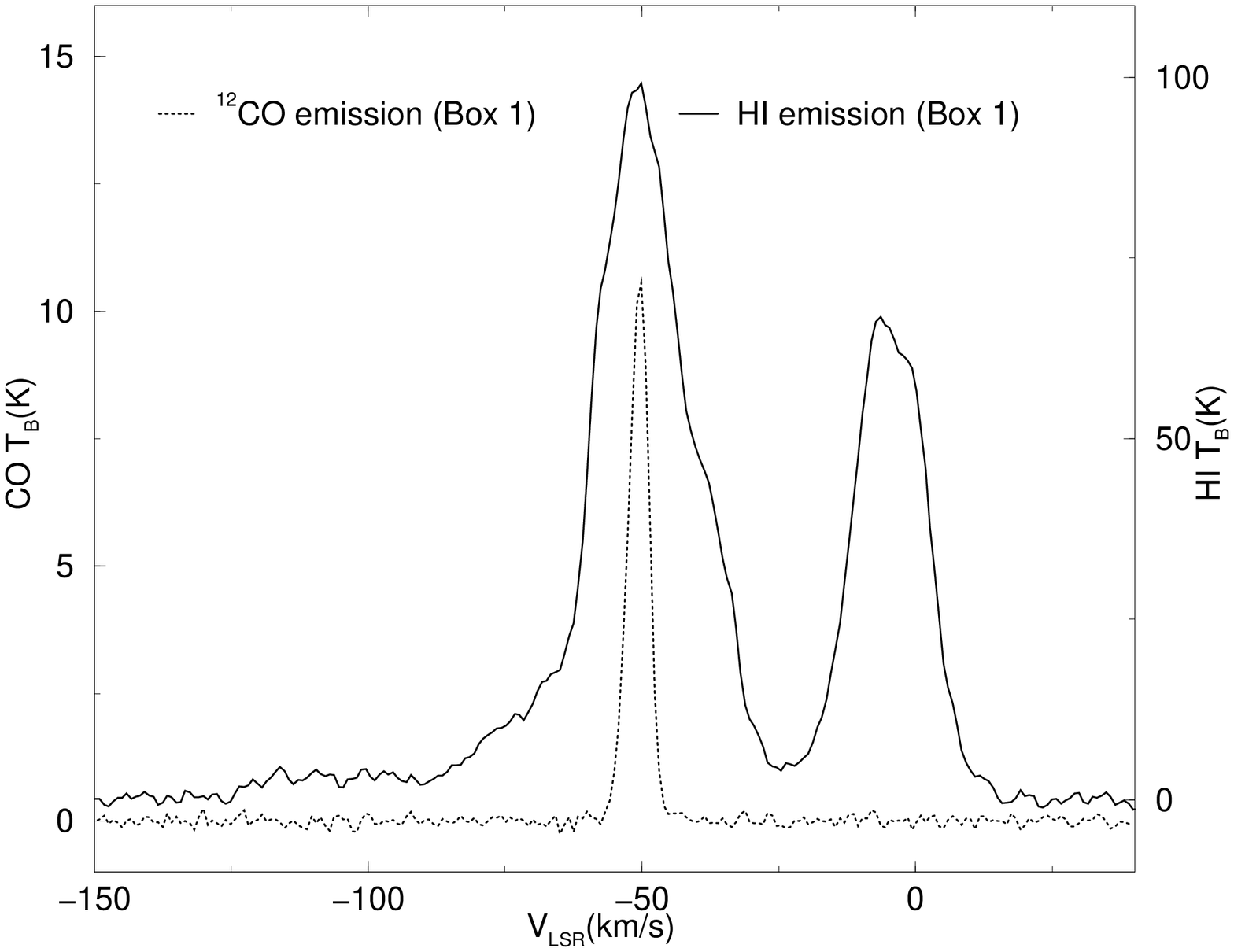}}
\put(0,230){\includegraphics{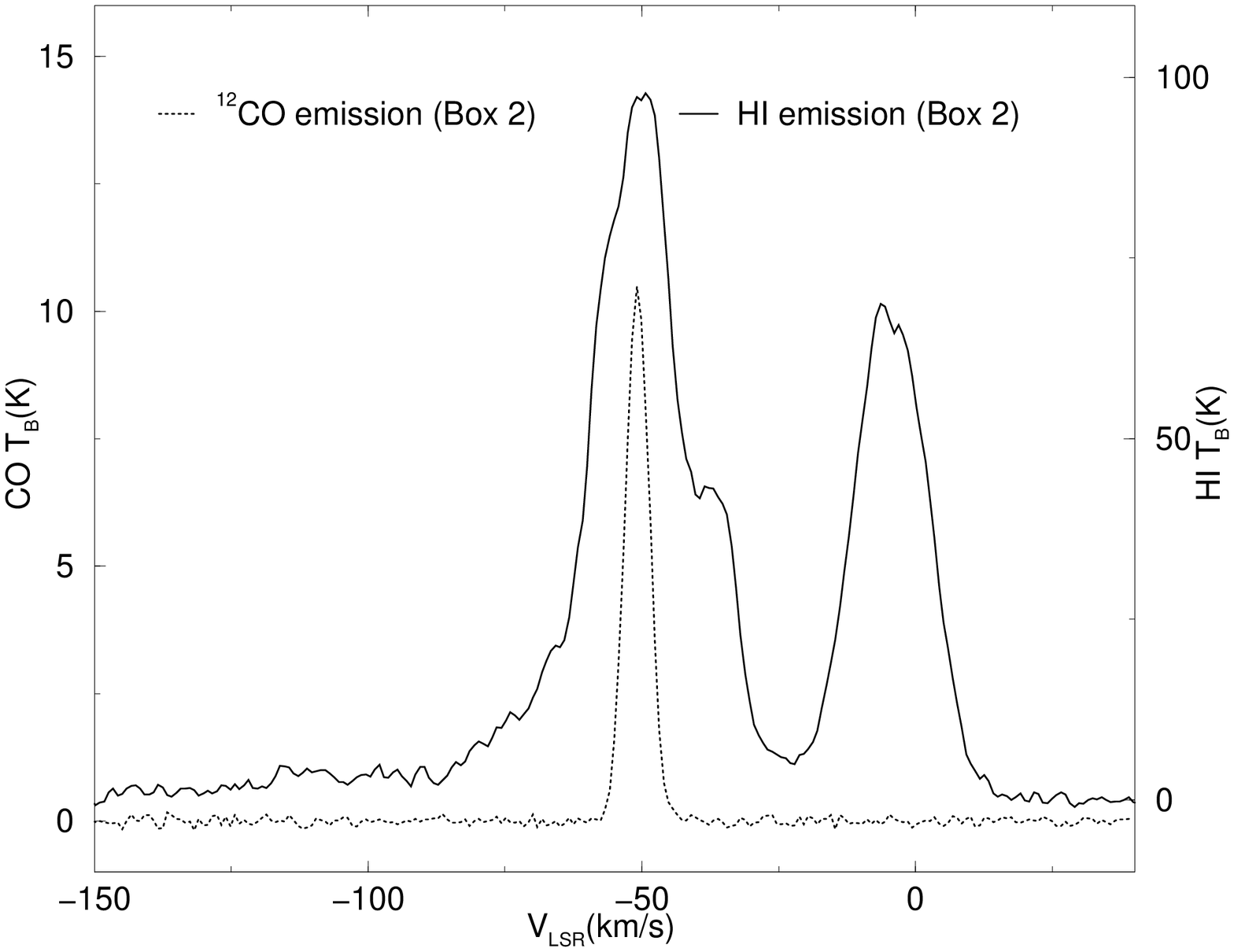}}
\put(0,110){\includegraphics{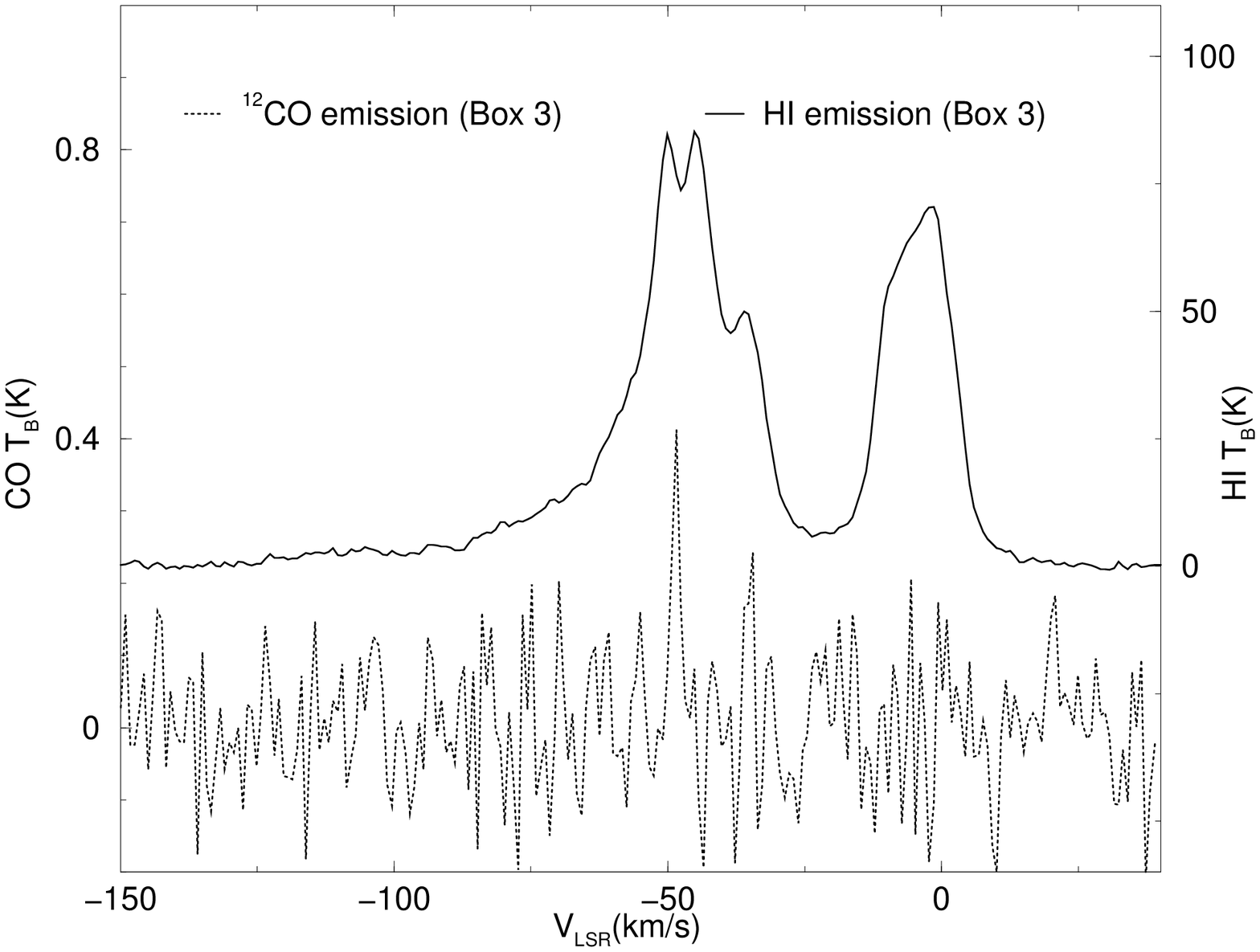}}
\put(0,-10){\includegraphics{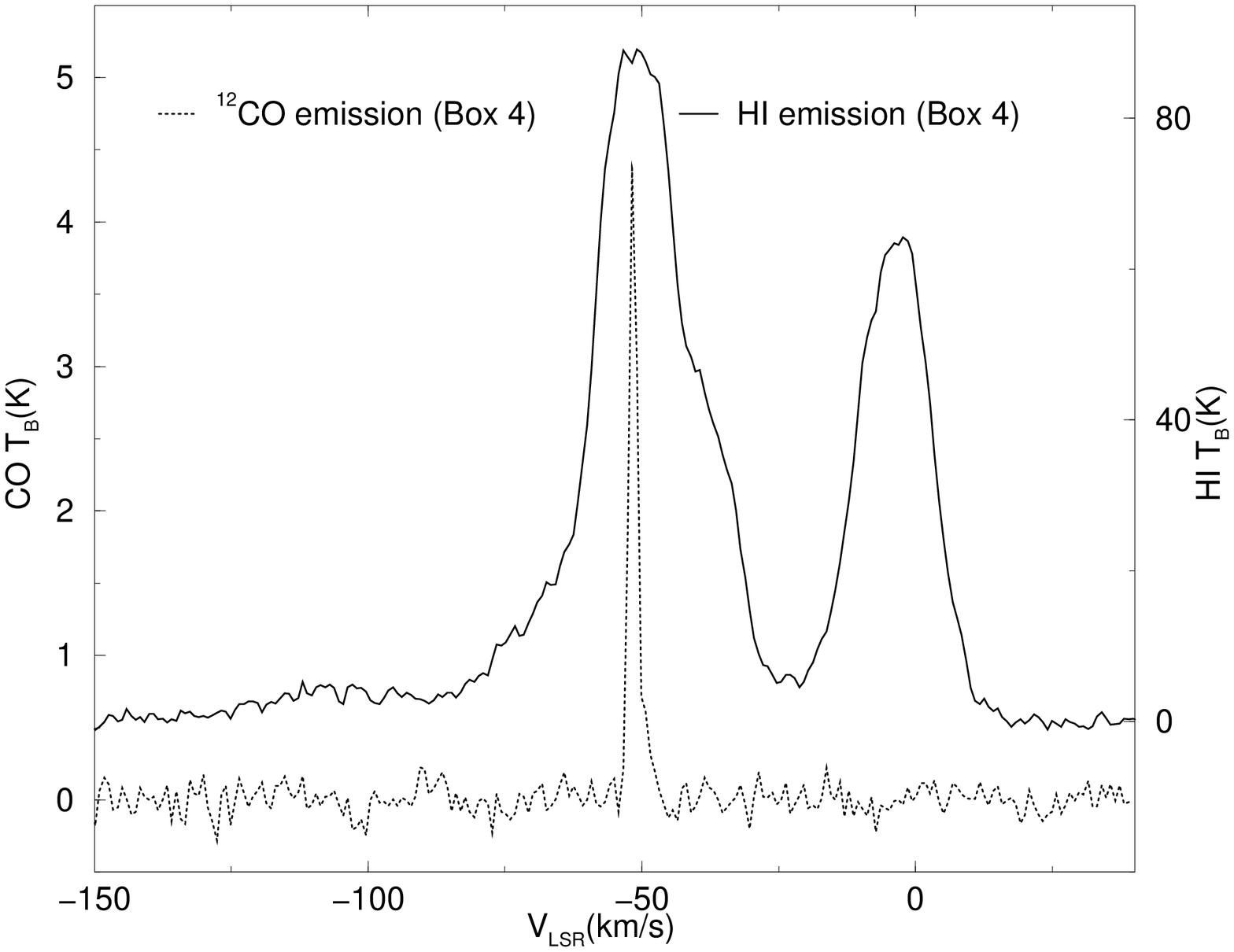}}
\end{picture}
\caption{HI and $^{12}$CO emission spectra from boxes 1 - 4}
\end{figure}

We searched the HI channel maps and HI emission spectra over the large region covered by the CO cloud complex, and found no HISA whose HI spectra show dips of $\ge$ 10 K with the FWHM $\Delta{v}$ of the absorption line 5 km s$^{-1}$. This is consistent with the non-detection of HISA features in the region in the CGPS HISA census \citep{Gibet05}. As an example, Fig. 3 shows the HI emission spectra and respective CO spectra extracted from boxes 1 to 4 shown in Fig. 1 (green). The top two rows of Fig. 3 clearly show no HISA features from the CO cloud at $-$51$\pm$5 km s$^{-1}$ against the warm HI background (T$_{HI}$$\sim$ 100K in the range of $-$40 to $-$60 km s$^{-1}$).
In contrast, a HISA feature from box 3 nearby CTB 109 is seen (the third row). 
We found one area of the CO cloud complex showing narrow CO lines and still bright enough CO emission (at $l$=108.85$^\circ$ and $b$=-1.02$^\circ$ that is labelled "4" in Fig. 1) to satisfy the conditions needed to produce HINSA. The CO emission spectrum from box 4 shown in the bottom of Fig. 3 has a narrow line width of 2 km s$^{-1}$. The respective HI emission spectrum shows no HINSA features.   
\section{Discussion and Conclusion}  
\subsection{The velocity/distance relation within the Perseus arm}
21 cm HI observations can provide distance estimates 
to Galactic objects using the radial velocity-distance relation based on the flat circular rotation curves model. However, this rotation model may lead to serious overestimation of an objects' distance, especially at the Perseus arm where the velocity field is strongly influenced by the spiral arm shock \citep{Rob72}. 
The velocity reversal inside the Perseus arm is caused by a spiral shock, while in other parts of the outer Galaxy the radial velocity decreases monotonically with distance. 
This velocity reversal creates a distance ambiguity for objects with radial velocities of $-$50 km s$^{-1}$ to $-$60 km s$^{-1}$ in the line-of-sight to CTB 109 \citep[TLF07 hereafter]{Tiaet07}. 

We use the two-arm HI model of \citet*{Foset06} (FM06 hereafter), which is an updated version of the \citet{Rob72} model. FM06 fit an empirical model of Galactic structure and density-wave motions to observations.
A comparison between photometric and the HI model distances to 22 objects (19 HII regions and 3 SNRs), shows individual point errors of 0.3 to 0.8 kpc (Fig. 8 of FM06).  The 20\% error quoted in FM06 is dominated by systematic errors.  The velocity of -55km s$^{-1}$ for CTB 109 gives 3.2 kpc in FM06 model if it is at the near side of the velocity reversal (Fig. 6 of TLF07). Since our new result is that CTB 109 is beyond the velocity reversal, the FM06 model gives 4.0 kpc distance. The difference in distance is 0.8 kpc. The error in this value is dominated by the systematic error of 0.16 kpc.

Because of the velocity reversal in the Perseus arm, HISA and HINSA should be produced if the cold CO clouds associated with CTB 109 are in front of the warm HI background, i.e. at distance of 3.2 kpc.
Absence of any HISA and HINSA features indicate that the cold CO cloud is at the far distance of 4 kpc in order to be behind the warm HI at the same velocity. We analyze this quantitatively below.  

\subsection{Is the amount of cold HI gas in the molecular cloud enough to produce significant HISA or HINSA against warm HI background?} 

First we consider the conditions for HISA. 
The intensity of HISA is given by $\Delta T_{HISA}$ = ($T_{B}(v)-T_{s}$)(1-$e^{-\tau}$).
With a normal background HI temperature of about 100k, to produce a measurable HISA signal
at about 10 K level \citep*{Andet09} requires an optical depth of HISA above 0.1.
Such optical depth is produced by a cold HI column of 1$\times$10$^{20}$ cm$^{-2}$ according to N$_{HI}$=1.82$\times$10$^{19}$$T_{s}$$\tau$$\Delta{v}$,
assuming the cold gas has a velocity dispersion of 5 km s$^{-1}$. The continuum brightness temperature of the southern bright spot of CTB109 is 30K, so the absorption feature at $-$51 km s$^{-1}$ should be produced by cold HI clouds at $-$51 km s$^{-1}$ (with $T_{s}$$\le$ 30K). From the absorption spectra, the column density of the HI absorbing clouds is 3$\times$10$^{19}$$T_{s}$ cm$^{-2}$, i.e. 3$\times$10$^{20}$ cm$^{-2}$ when $T_{s}$=10 K. This is more than enough to produce measurable HISA if the cloud is in front of the warm HI emission (100 K).   

Next we calculate the optical depth of an HI absorption line which is required to produce a HINSA feature with peak intensity of 3 $\sigma$. The rms of the HI emission data is about 1.7 K per channel calculated from the HI spectrum of G108.77-0.95 (Fig. 2). \citet{Tayet03} give that the rms noise level in brightness temperature for an empty channel ranges from 2.1 K to 3.2 K throughout the survey. 
We take the value of 2.1 K here. Based on a three component radiative transfer calculation and the assumption that the HI emission is optically thin, we have  (Equation 8 of \citet*{Liet03},
\begin{equation}
T_{ab}= [pT_{HI} + T_c - T_x](1-e^{-\tau}) \>\> ,
\end{equation}
where the depth of absorption $T_{ab}$ is determined by the fraction of HI in the background $p$, the emission temperature $T_{HI}$, the continuum background $T_c$, and the cold HI excitation temperature $T_x$. Assuming a uniform Galactic HI disk with a 17 kpc radius, it is estimated that $p=0.815$ toward the direction of CTB 109. Using the reasonable assumption of $T_c = 3.5 K$ (including CMB and interstellar radiation field see \citet*{Liet03}),  $T_{HI}=100 K$ and $T_x = 10 K$, an optically depth of $\tau = 0.084$ is required to produce an absorption line of 6.3 K peak depth (3 $\sigma$). For a FWHM line-width of 2.0 km s$^{-1}$ of HINSA, this corresponds to an upper limit of the column density of cold HI $N_{HI}$ = 3.2$\times10^{18}$ cm$^{-2}$.
Theoretically, the abundance of atomic HI inside molecular clouds is in the range of 0.05 to 0.27\% \citep*{Golet05}. Observationally, using an improved technique to measure the ratio, \citet{Krcet08} obtained a precise value for two clouds, i.e. 0.09\% for L134 and 0.53\% for L1757. Taking the theoretical mean HI/H$_{2}$ ratio of 0.16\%, the non-detection of HINSA toward the large molecular cloud interacting with CTB 109 puts a 3$\sigma$ upper limit on the cloud $H_2$ column density at 2.0$\times 10^{21}$cm$^{-2}$.  

The $^{12}$CO spectrum from box 4 (Fig. 3) gives W$_{^{12}CO}$ = 12 $K km s^{-1}$ for the cloud. The $^{12}$CO to H$_{2}$ conversion factor $X$ is in the range of 2.2 to 2.8$\times$ 10$^{20} [cm^{-2}/(K km s^{-1}$)] based on three calibration techniques \citep*{Solet91}. This gives an H$_{2}$ column density of 3 $\times$10$^{21}$ cm$^{-2}$ for this cloud ($X$ = 2.5$\times$10$^{20}$ cm$^{-2}/(K km s^{-1}$)). This is enough to produce measurable HINSA at 4.5$\sigma$ level if the cloud is in front of the warm HI at the same velocity.

Fig. 3 (third row) shows a good example that a HISA feature at $-$50 km s$^{-1}$ is clearly detected nearby CTB 109. The CO cloud at $-$50 km s$^{-1}$ has low brightness temperature of 0.3 K and shows no physical relation with the remnant. The W$_{^{12}CO}$ from the cloud is about 1 K km/s so it has an H$_{2}$ column density of 2.5 $\times$10$^{20} cm^{-2}$. This is about 30 times lower than that from the cloud complex associated with Sh 152. This indicates that the cold HI cloud at $-$50 km s$^{-1}$ is likely responsible for the absorption feature.       
\subsection{Distance of 4 kpc to CTB 109/AXP 1E 2259+586} 
The non-detection of both HISA and HINSA leads to the conclusion that the cloud at $-$55 km s$^{-1}$ is behind the warm HI and thus CTB 109, interacting with the cloud, is behind the velocity reversal in the Perseus arm. 
As noted above, the HI model distance of 4 kpc has a uncertainty of 20\% (i.e. 0.8 kpc). \citet*{Braet93} have studied the observed velocity field of the outer galaxy and found residuals from circular motions. They found that these velocity residuals are consistent with streams motions due to spiral arms, supporting the Roberts (and thus FM06)  models.

K02 estimated a distance for CTB 109 of 3$\pm$0.5 kpc based on two arguments. First, the HI absorption at $-$45 km s$^{-1}$ in the extragalactic source is not seen in the spectra of CTB 109 or Sh 152, so they argued CTB 109 and Sh 152 must be at the nearer distance in the spiral shock region. However, the extragalactic source is not near to CTB 109 (outside of the area shown in Fig.1) and thus spatial variations in the HI distribution can result in lack of significant HI at $-$45 km s$^{-1}$ in front of CTB 109 and invalidate that conclusion.
Secondly they considered the spectroscopic distances and radial velocities of 11 HII regions which in the Perseus arm nearby on the sky to CTB 109. These HII regions have an average distances of 3 kpc. However the two HII regions, Sh 152 ([108.77,-0.95]) and Sh 149 ([108.34, -1.12]), closest to CTB 109 on the sky have spectroscopic distances of 3.6$\pm$1.1 kpc and 5.4$\pm$1.7 kpc, respectively. This does not favour the small distance to CTB 109 of 3 kpc. These two HII regions have respective radial velocities of $-$50.4$\pm$0.5 km s$^{-1}$ (or $-$49.1$\pm$0.9 km s$^{-1}$ from \citet{Loc89} and $-$53.1$\pm$1.3 km s$^{-1}$ from \citet*{Braet93} which are similar to the radial velocity ($-$55 km s$^{-1}$) of the molecular cloud complex with which CTB 109 is interacting. This also support the larger distance for CTB 109 of 4.0 kpc. 

CTB 109 is host to AXP 1E 2259+586. DvK06 excluded small distance of 3 kpc for 1E 2259+586 due to absence of red clump stars sufficiently highly reddened to be consistent with 1E 2259+589's column density. There are two jumps (at $A_{V}$ 3 and 8) in their Fig. 7 which shows a redding--distance relation along the line-of-sight to AXP 1E 2259+586. They favour a large distance of 6 kpc for the AXP based on the large extinction of 8. The HI absorption spectrum of CTB 109 (Fig. 2) excludes a large distance. Eg for 6 kpc the radial velocity of FM06 model is about $-$80 km s$^{-1}$, but no absorption is seen beyond $-$56km s$^{-1}$. Our distance measurement of 4.0 kpc to AXP 1E 2259+586 puts the AXP at the far edge of the Perseus arm and would give a small extinction of $A_{V} \sim$ 3 for the AXP. The reddening stays at $A_{V} \sim$ 3 out to about 6 kpc (Fig. 7 of DvK06), implying excess local extinction for the AXP at 4 kpc. Using N$_{H}$ =  $A_{V}$ $\times$ 2.3$\times$10$^{21}$ cm$^{-2}$, with inferred excess $\Delta{A_{V}}$ = 5, gives excess  $\Delta{N_{H}}$ $\approx$ 10$^{22}$ cm$^{-2}$. This is quite plausible given that AXPs should be produced in SN explosions of massive stars and these SN also are known to produce large amount of dust in their ejecta. 

The new distance of 4 kpc to the AXP leads to a larger explosion energy of 1.3$\times$10$^{51}$ ergs than that based on 3 kpc \citep*{Vinet06}. However this is not a large enough increase to argue that CTB 109 had an unusually large explosion energy. The same situation occurs for SNR Kes 73/AXP 1E 1841-045 system \citep*{Tiaet08b}.  Both cases have larger distances than previously estimated but still give normal explosion energies (10$^{51}$ ergs).  This adds support to \citet*{Vinet06}'s argument against the possibility that the magnetars are formed from rapidly rotating (less than a few millisecond) proto-neutron stars.  We have applied a new technique for determining the distance to CTB 109, i.e. absence or presence of HISA and HINSA. This is a new useful tool for finding distances to other objects in the Galactic plane. 

\section{Acknowledgments}
WWT acknowledges support of the CBP and the NSFC. DAL thanks support from the Natural Sciences and Engineering Research Council of Canada. Di Li's work is supported by the Jet Propulsion Laboratory, California Institute of Technology, under a contract with the National Aeronautics and Space Administration.  The DRAO is operated as a national facility by the National Research Council of Canada.

\end{document}